\documentclass[12pt]{article}
\usepackage{a4wide}
\usepackage{graphics}

\makeatletter

\newcommand{\LyX}{L\kern-.1667em\lower.25em\hbox{Y}\kern-.125emX\@}

\makeatother

\begin{document}

\title{Nucleation theory and the phase diagram of the magnetization-reversal transition}

\maketitle
{\par\centering Arkajyoti Misra\footnote{
e-mail : arko@cmp.saha.ernet.in
} and Bikas K. Chakrabarti\footnote{
e-mail : bikas@cmp.saha.ernet.in
} \par}

{\par\centering \emph{Saha Institute of Nuclear Physics, 1/AF Bidhannagar, Calcutta
700 064, India.}\par}




\begin{abstract}
The phase diagram of the dynamic magnetization-reversal transition in pure Ising
systems under a pulsed field competing with the existing order can be explained
satisfactorily using the classical nucleation theory. Indications of single-domain
and multi-domain nucleation and of the corresponding changes in the nucleation
rates are clearly observed. The nature of the second time scale of relaxation,
apart from the field driven nucleation time, and the origin of its unusual large
values at the phase boundary are explained from the disappearing tendency of
kinks on the domain wall surfaces after the withdrawal of the pulse. The possibility
of scaling behaviour in the multi-domain regime is identified and compared with
the earlier observations.\newpage

\end{abstract}
The study of dynamical phase transitions in pure Ising systems, with ferromagnetic
short range interactions, under the influence of time dependent external magnetic
field has recently become one of the significant areas of interest in statistical
physics \cite{acrmp}\cite{srnpre}\cite{mcpre}. The effect of external magnetic
fields which are periodic in time was first dealt with mean field theory \cite{to}.
Subsequently, through extensive Monte Carlo studies, the existence of a dynamic
phase transition was established and properly characterized \cite{srnpre}\cite{acprb}\cite{srnprl}.
A relevant investigation in this context was the study of the effect of external
pulsed field which is uniform in space but applied for a finite duration. All
these pulsed field studies are concerned with the system below its static critical
temperature \( T_{c}^{0} \), where the system has a long range order characterized
by a non-zero magnetization \( m_{0} \). Under the influence of a `positive'
field or a field applied along the direction of prevalent order, the system
does not show any new phase transition \cite{abc}. However, a `negative' pulsed
field competes with the existing order and the system may show a transition
from an initial equilibrium state with magnetization \( m_{0} \) to a final
equilibrium state with magnetization \( -m_{0} \) \cite{mcpa}. Such a transition
can be brought about by tuning the duration (\( \Delta t \)) and strength (\( h_{p} \))
of the pulse, and the `phase diagram' in the \( h_{p}-\Delta t \) plane gives
the minimal combination of these two parameters to bring about the transition
at any temperature \( T \) below \( T_{c}^{0} \). The transition is dynamic
in nature and both length and time scales are seen to diverge across this transition
\cite{mcpre}\cite{smc}. 

According to classical theory of nucleation, there could be two mechanisms through
which the droplet of a particular spin grows in the sea of opposite spins. When
the external magnetic field is relatively weak, a single droplet grow to cover
the whole system size and this regime is called the single-droplet (SD) \cite{srnprl}
or nucleation regime \cite{as}. On the other hand, under stronger magnetic
fields, many small droplets grow simultaneously and eventually they coalesce
to span the entire system. This is called the multi-droplet (MD) or coalescence
regime\cite{srnprl}\cite{as}. The crossover from the SD to the MD regime takes
place at the dynamic spinodal field or \( H_{DSP} \), which is a function of
the system size \( L \) and of the temperature \( T \). There are two time
scales in the problem : (i) the nucleation time \( \tau _{N} \) is the time
taken by the system to leave the metastable state under the influence of the
external field and (ii) relaxation time \( \tau _{R} \) which is defined as
the time taken by the system to reach the final equilibrium state after the
external field is withdrawn. In this letter, we show that the phase diagram
of the magnetization-reversal transition can be explained satisfactorily by
employing the classical nucleation theory \cite{gd}. The nature of the transition
changes form a discontinuous to a continuous one, giving rise to a `tricritical'
point as the system goes from the SD regime to the MD regime, depending on the
temperature and the strength of the applied pulse. The dimension dependent factor
by which the nucleation time differs in the two regimes is confirmed by Monte
Carlo simulation of \( 2d \) kinetic Ising model evolving under Glauber dynamics
below \( T^{0}_{c} \). Our observations here also support the validity of the
finite size scaling of the of the order parameter used earlier to obtain the
dynamic critical exponents of the transition \cite{mcpre}.

Monte Carlo simulation using single spin-flip Glauber dynamics have been used
to study pure Ising system in 2\( d \) below \( T_{c}^{0} \) (\( \simeq 2.27,\textrm{ } \)
in units of the strength of the cooperative interaction). The system is subjected
to a magnetic field \( h(t) \) of finite strength \( h_{p} \) for a finite
duration \( \Delta t \) :

\begin{equation}
\label{eqn1}
\begin{array}{ccrlc}
h(t) & = & -h_{p} & \textrm{for }t_{0}\leq t\leq t_{0}+\Delta t & \\
 & = & 0 & \textrm{otherwise}.
\end{array}
\end{equation}
In order to prepare the system in an ordered state corresponding to a temperature
\( T<T_{c}^{0} \), \( t_{0} \) is taken much larger than the equilibrium relaxation
time of the system at that temperature. The negative sign signifies that the
applied field competes with the prevalent order characterized by equilibrium
magnetization \( m_{0} \), which is a function of \( T \) only. Depending
on the values of \( h_{p} \) and \( \Delta t \), the system can either go
back to the original state (with magnetization \( m_{0} \)) or to the other
equivalent ordered state, characterized by equilibrium magnetization \( -m_{0} \),
eventually after the withdrawal of the field. This gives rise to the magnetization-reversal
transition which brings the system from one equilibrium ordered state to the
other, below \( T_{c}^{0} \). 

It appears that the average magnetization at the time of withdrawal of the pulse
\( m_{w}\equiv m(t_{0}+\Delta t) \) is a very relevant quantity : depending
on the sign of \( m_{w} \), the system chooses the final equilibrium state.
On an average, the transition to the \( -m_{0} \) state occurs for negative
values of \( m_{w} \), whereas for positive \( m_{w} \) the system goes back
to the original \( +m_{0} \) state. There occur of course fluctuations, where
the transition to the opposite order takes place even for small positive \( m_{w} \)
values or vice versa. The phase boundary is therefore identified as the \( m_{w}=0 \)
line. Fig. 1 shows the Monte Carlo phase boundaries at a few different temperatures
below \( T_{c}^{0} \) for a 100\( \times  \)100 lattice with periodic boundary
condition. At very low temperatures (typically for \( T<0.5 \)), the transition
is discontinuous in nature all along the phase boundary. Crossover to a continuous
transition region appears along the phase boundary for higher temperatures,
thereby giving rise to a tricritical point. Along a typical phase boundary where
both kinds of transition are observed, the continuous transition region appears
for smaller values of \( \Delta t \) (higher values of \( h_{p} \)), whereas
the discontinuous region appears for higher values of \( \Delta t \) (smaller
values of \( h_{p} \)). 

It is instructive to look at the droplet picture of the classical nucleation
theory to describe the nature of the phase diagram for the magnetization-reversal
transition. The typical configuration of a ferromagnet below \( T_{c}^{0} \)
consists of clusters or droplets of down spins in a background of up spins or
vice versa. According to the classical nucleation theory \cite{gd}, the equilibrium
number of droplets of size \( l \) is then given by \( n_{l}=N\exp (-\beta \epsilon _{l}) \),
where \( \beta =1/k_{B}T \) and \( \epsilon _{l} \) is the free energy of
formation of a droplet of size \( l \). Assuming spherical shape of the droplets,
one can write \( \epsilon _{l}=2hl+\sigma (T)l^{(d-1)/d} \), where \( \sigma  \)
is the temperature dependent surface tension and \( h \) is the external magnetic
field. Droplets of size greater than a critical value \( l_{c} \) are then
favoured to grow. One obtains \( l_{c} \) by maximizing \( \epsilon _{l} \)
: \( l_{c}=\left[ \sigma (d-1)/2d\left| h\right| \right] ^{d} \). The number
of supercritical droplets is then given by \( n_{l}^{*}=N\exp \left( -\beta K_{d}\sigma ^{d}/\left| h\right| ^{d-1}\right)  \),
where \( K_{d} \) is a constant depending on dimension only. In the nucleation
regime or the SD regime, there is only one supercritical droplet that grows
and engulfs the entire system. The nucleation time \( \tau ^{SD}_{N} \) is
inversely proportional to the nucleation rate \( I \). According to the Becker-D\( \ddot{\textrm{o}} \)ring
theory \cite{gd}, \( I \) in turn is proportional to \( n^{*}_{l} \), and
thus

\[
\tau ^{SD}_{N}\propto I^{-1}\propto \exp \left( \frac{\beta K_{d}\sigma ^{d}}{\left| h\right| ^{d-1}}\right) .\]
 In the coalescence or MD regime, where due to stronger magnetic field many
droplets grow simultaneously and eventually coalesce to form a system spanning
droplet, the nucleation time is given by \cite{as} 
\[
\tau ^{MD}_{N}\propto I^{-1/(d+1)}\propto \exp \left( \frac{\beta K_{d}\sigma ^{d}}{(d+1)\left| h\right| ^{d-1}}\right) .\]
During the time when the field is `on', the only relevant time scale of the
problem is the nucleation time. The phase boundary of the magnetization-reversal
transition corresponds to the threshold value of the field pulse (\( h_{p}^{c} \)),
which can bring the system from an equilibrium state with magnetization \( m_{0} \)
to a state with magnetization \( m_{w}=0_{-} \) in time \( \Delta t \), so
that the system eventually evolves to a state with magnetization \( -m_{0} \).
Equating therefore \( \Delta t \) with the nucleation time \( \tau _{N} \),
one gets for the magnetization-reversal phase diagram 
\begin{equation}
\label{eqn2}
\begin{array}{cccccl}
\ln (\Delta t) & = & c_{1} & + & \frac{C}{h_{p}^{d-1}} & \textrm{in SD regime}\\
 & = & c_{2} & + & \frac{C}{(d+1)h_{p}^{d-1}} & \textrm{in MD regime},
\end{array}
\end{equation}
where \( C=\beta K_{d}\sigma ^{d} \) and \( c_{1} \), \( c_{2} \) are constants.
Therefore a plot of \( \ln (\Delta t) \) against \( \left( h_{p}^{c}\right) ^{d-1} \)
would show different slopes in the two regimes. This is indeed the case as shown
in Fig. 2, where two distinct slopes are seen at higher temperatures (typically
for \( T>0.5 \)). The ratio of the slopes \( R \) corresponding to the two
regimes has a value close to \( 3 \) as predicted for \( 2d \) by the classical
nucleation theory. The point of intersection of the two straight lines gives
the value of \( H_{DSP} \) at that temperature and system size. At lower temperatures,
however, the MD region is absent which is marked by a single slope (Fig. 2(d)).

According to the classical nucleation theory \( l_{c}\propto \left| h_{p}\right| ^{-d} \)
and at any fixed temperature, therefore, one expects stronger (weaker) fields
to bring the system to the MD (SD) regime. Fig. 3 shows snapshots of the spin
configuration at a particular temperature, where the dots correspond to down
spins. It is clear from the figure that many droplets of down spins are formed
for a smaller value of \( \Delta t \) or equivalently for large \( h_{p} \)
whereas only a single down spin droplet is formed for a larger \( \Delta t \)
or weaker \( h_{p} \). The snapshots of the system are taken at the time of
withdrawal of the pulse, beyond which the system is left to itself to relax
back to either of the equilibrium states. The time taken by the system to reach
the final equilibrium after the withdrawal of the pulse is defined as the relaxation
time (\( \tau _{R} \)). It is observed \cite{mcpre} that \( \tau _{R}\sim \kappa (T,L)\exp \left[ -\lambda (T)\left| m_{w}\right| \right]  \),
where \( \kappa (T,L) \) is a constant depending on temperature and system
size (\( \kappa \rightarrow \infty \textrm{ as }L\rightarrow \infty  \)) and
\( \lambda (T) \) is a constant depending on temperature. This is in distinct
contrast with the normal relaxation of a ferromagnet to the equilibrium state
with magnetization \( m_{0} \), starting from a random initial state where
the average magnetization is close to zero. The effect of the field is to initiate
the nucleation process and by the time the pulse is withdrawn, the droplet(s)
has (have) very few kinks along the surface(s). Once a droplet or a domain forms
a flat boundary, it becomes a rather stable configuration and the domain wall
movement practically stops; thereby restricting further nucleation. When the
system is trapped in such a metastable state, it is left for large fluctuations
to initiate further movement of the domain walls and resume the process of nucleation.
The closer to zero is \( m_{w} \), more is the chance that the system will
be trapped in a metastable state owing to more number of flat domain walls.
Thus the effect of the pulse is to create a domain structure which does not
favour fast nucleation and even for \( T\ll T_{c}^{0} \), \( \tau _{R} \)
therefore diverges at the phase boundary.

To determine the order of the phase transition one can look at the probability
distribution \( P(m_{w}) \) of \( m_{w} \), as one approaches a phase boundary.
Fig. 4 shows the variation of \( P(m_{w}) \) across the phase boundary corresponding
to a particular \( T \) at two different values of \( \Delta t \). The existence
of a single peak of \( P(m_{w}) \) in (a), which shifts its position continuously
from \( +1 \) to \( -1 \), indicates the continuous nature of the transition;
whereas in (b), one obtains two peaks simultaneously, close to \( \pm m_{0}(T) \),
indicating that the system can simultaneously reside in two different phases
with finite probabilities. This is a sure indication of discontinuous phase
transition. The crossover from the discontinuous transition to a continuous
one along the phase boundary is not very sharp; instead one gets a `tricritical
region' where the determination of the nature of the transition is not very
conclusive. As is evident from Fig. 2, the data points in this particular region
do not fit to either of the two straight lines corresponding to the two regimes.
Thus for \( h_{p}^{c}\ll H_{DSP} \) the system belongs to the MD regime and
the nature of the transition is continuous, whereas for \( h_{p}^{c}\gg H_{DSP} \)
the system is brought to the SD regime where the transition is discontinuous
in nature.

In our earlier work \cite{mcpre}, we clearly noticed that finite size scaling
of the order parameter was possible only at higher temperatures for moderately
low \( \Delta t \). The reason behind that becomes clear now, as at higher
temperatures the tricritical point shifts towards higher values of \( \Delta t \)
and hence in the lower \( \Delta t \) region the system belongs to the MD regime
and the transition is continuous in nature. This supports the scaling. Since
at lower temperatures the continuous region gradually shrinks before disappearing
altogether, all attempts for a finite size scaling fit failed. This observation
also compares with that of Sides et al. \cite{srnprl} for the scaling behaviour
in the case of dynamic transition under periodic fields.

In this letter we have shown that the phase diagram of the magnetization-reversal
transition in the Ising model can be explained satisfactorily from the classical
nucleation theory. Across \( H_{DSP} \), the system goes from MD or coalescence
regime where the transition is continuous in nature to SD or nucleation regime
where the transition is discontinuous. A tricritical point separates these two
regimes which moves towards smaller \( \Delta t \) (larger \( h_{p} \)) values
as one decreases \( T \) until it disappears altogether at low temperatures.
The dimensional factor in the nucleation time of the two regimes is found to
be quite accurately reproduced in the Monte Carlo simulations. There are two
time scales in the problem, viz. \( \tau _{N} \) and \( \tau _{R} \). \( \tau _{N} \),
which is determined by the pulse duration \( \Delta t \), brings the system
from one regime to the other depending on the temperature; while \( \tau _{R} \)
determines the relaxation of the field-released system and it diverges at the
phase boundary even for \( T\ll T_{c}^{0} \). Finite size scaling of the order
parameter for the transition is also justified in the MD regime where the transition
is continuous in nature.

{\par\centering {*}{*}{*}\par}

A. M. would like to thank Muktish Acharyya and Burkhard Duenweg for useful discussions. 

\textbf{\large Figure Captions}{\large \par}

\noindent Fig. 1. Monte Carlo phase diagram for magnetization-reversal transition
in a \( 100\times 100 \) lattice: (a) \( T=0.2 \), (b) \( T=0.6 \), (c) \( T=1.0 \),
(d) \( T=1.4 \), (e) \( T=2.0 \).

\noindent Fig. 2. Plot of \( \ln (\Delta t) \) against \( \left[ h_{p}^{c}(\Delta t,T)\right] ^{-1} \),
obtained from the phase diagram in Fig. 1, at different temperatures. (a) \( T=1.4 \),
(b) \( T=1.0 \), (c) \( T=0.7 \), (d) \( T=0.2 \). The observed values of
the slope ratio \( R \) are close to \( 3.38 \), \( 3.09 \) and \( 3.24 \)
in (a), (b) and (c) respectively.

\noindent Fig. 3. Snapshots of the spin configuration of the system at \( t=t_{0}+\Delta t \)
for \( T=1.0 \); (a) \( \Delta t=2 \), \( h_{p}=1.91 \) (MD regime) (b) \( \Delta t=100 \),
\( h_{p}=0.65 \) (SD regime).

\noindent Fig. 4. Probability distribution \( P(m_{w}) \) of \( m_{w} \) at
\( T=0.6 \). In (a) where \( \Delta t=1 \) (SD regime) , the position of the
peak changes sharply from \( +1 \) to \( -1 \) indicating a discontinuous
transition: (i) \( h_{p}=2.05 \), (ii) \( h_{p}=2.15 \), (iii) \( h_{p}=2.25 \).
In (b) where \( \Delta t=500 \) (MD regime) , the position of the peak changes
continuously from \( +1 \) to \( -1 \) indicating a continuous transition:
(i) \( h_{p}=0.92 \), (ii) \( h_{p}=1.02 \), (iii) \( h_{p}=1.12 \).
\newpage

\vspace{0.3cm}
{\par\centering \includegraphics{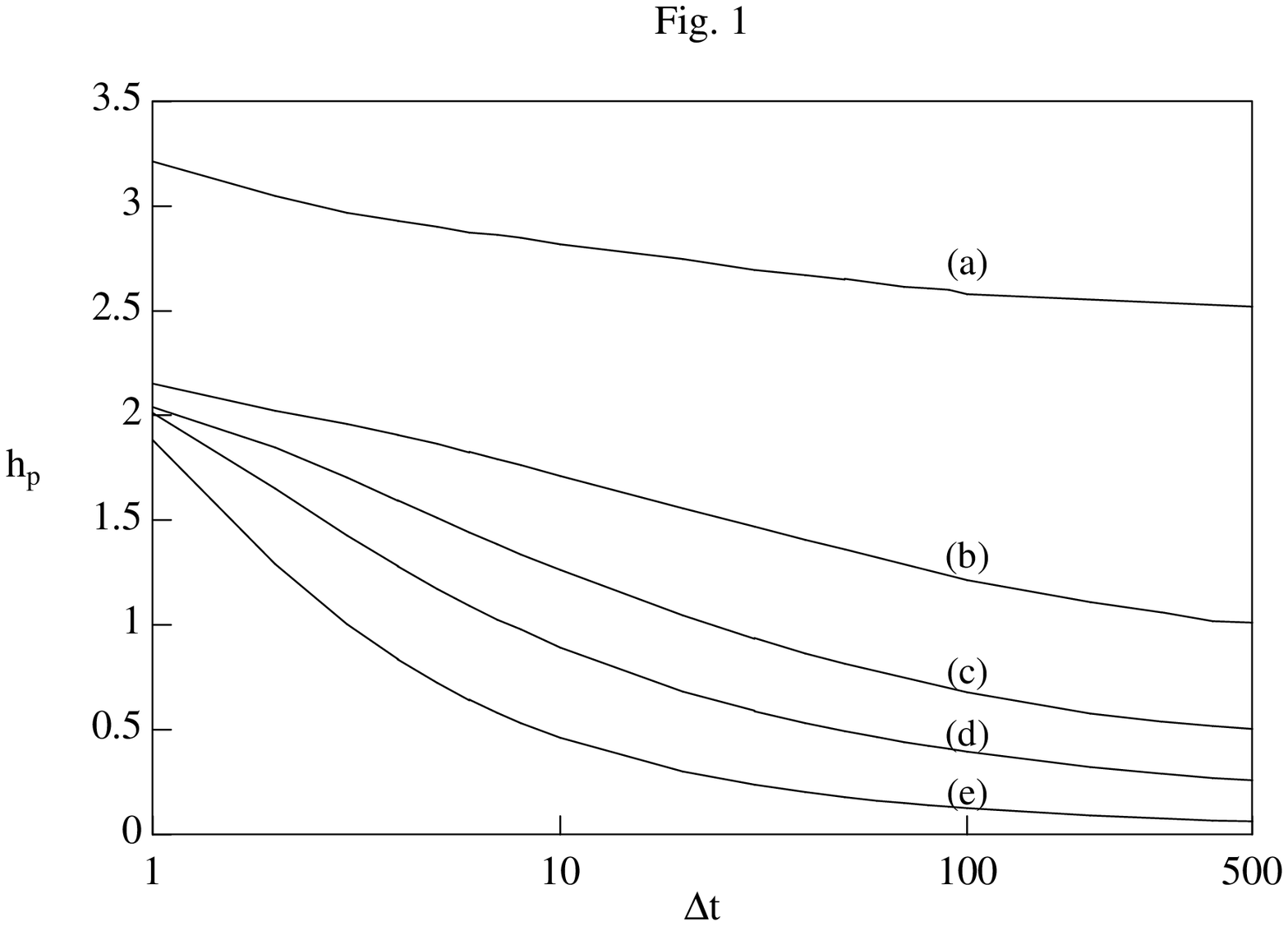} \par}
\vspace{0.3cm}\newpage

\vspace{0.3cm}
{\par\centering \includegraphics{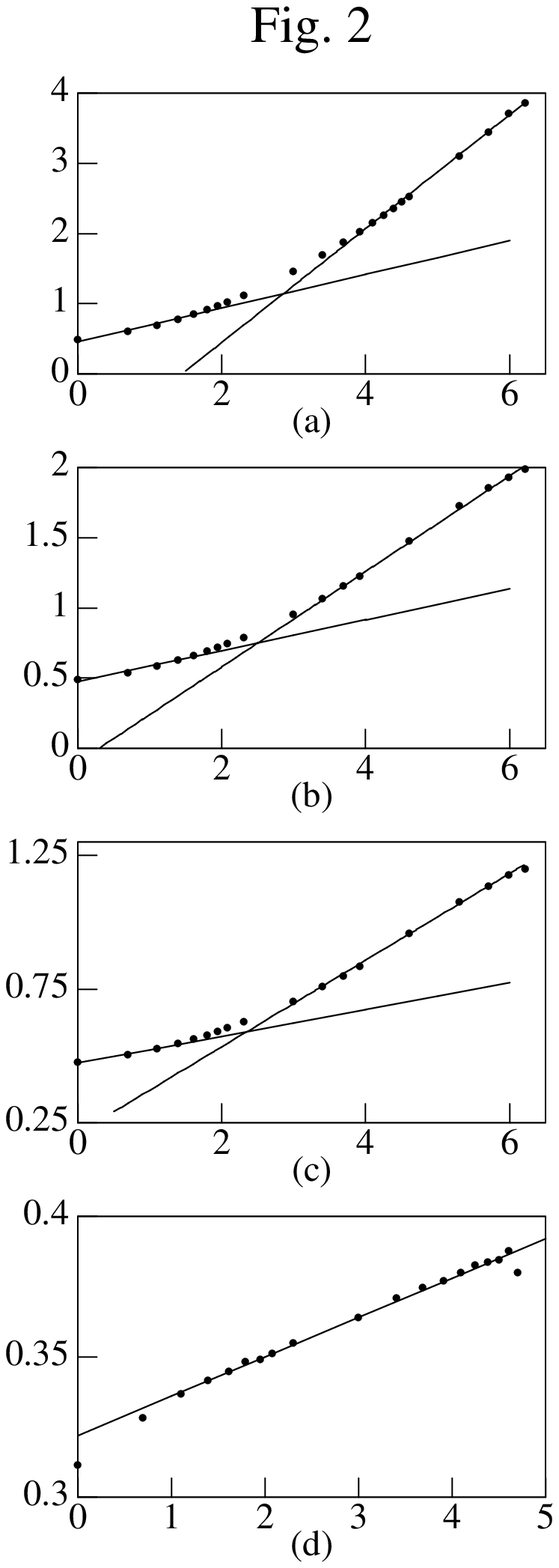} \par}
\vspace{0.3cm}\newpage

\vspace{0.3cm}
{\par\centering \includegraphics{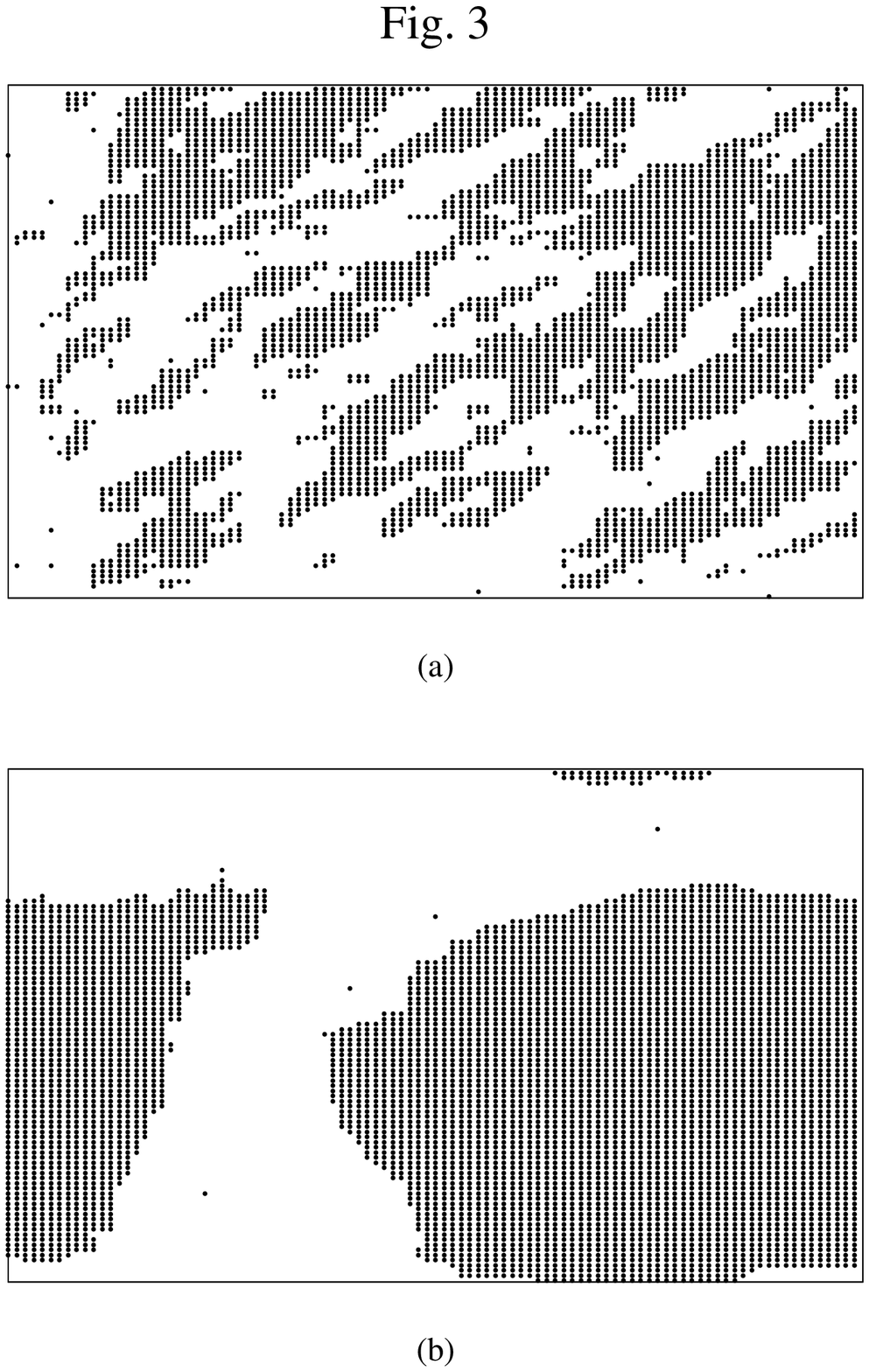} \par}
\vspace{0.3cm}\newpage

\vspace{0.3cm}
{\par\centering \includegraphics{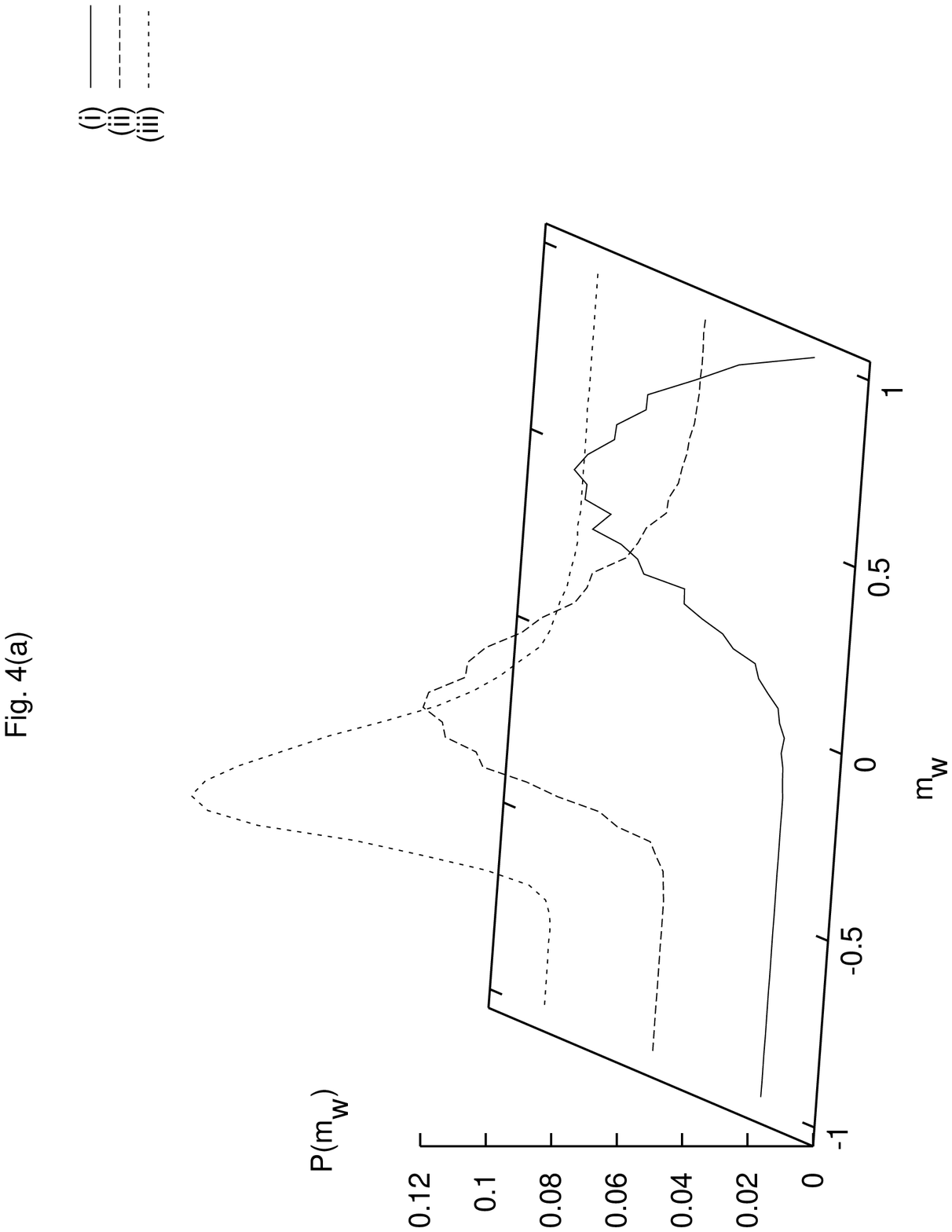} \par}
\vspace{0.3cm}\newpage

\vspace{0.3cm}
{\par\centering \includegraphics{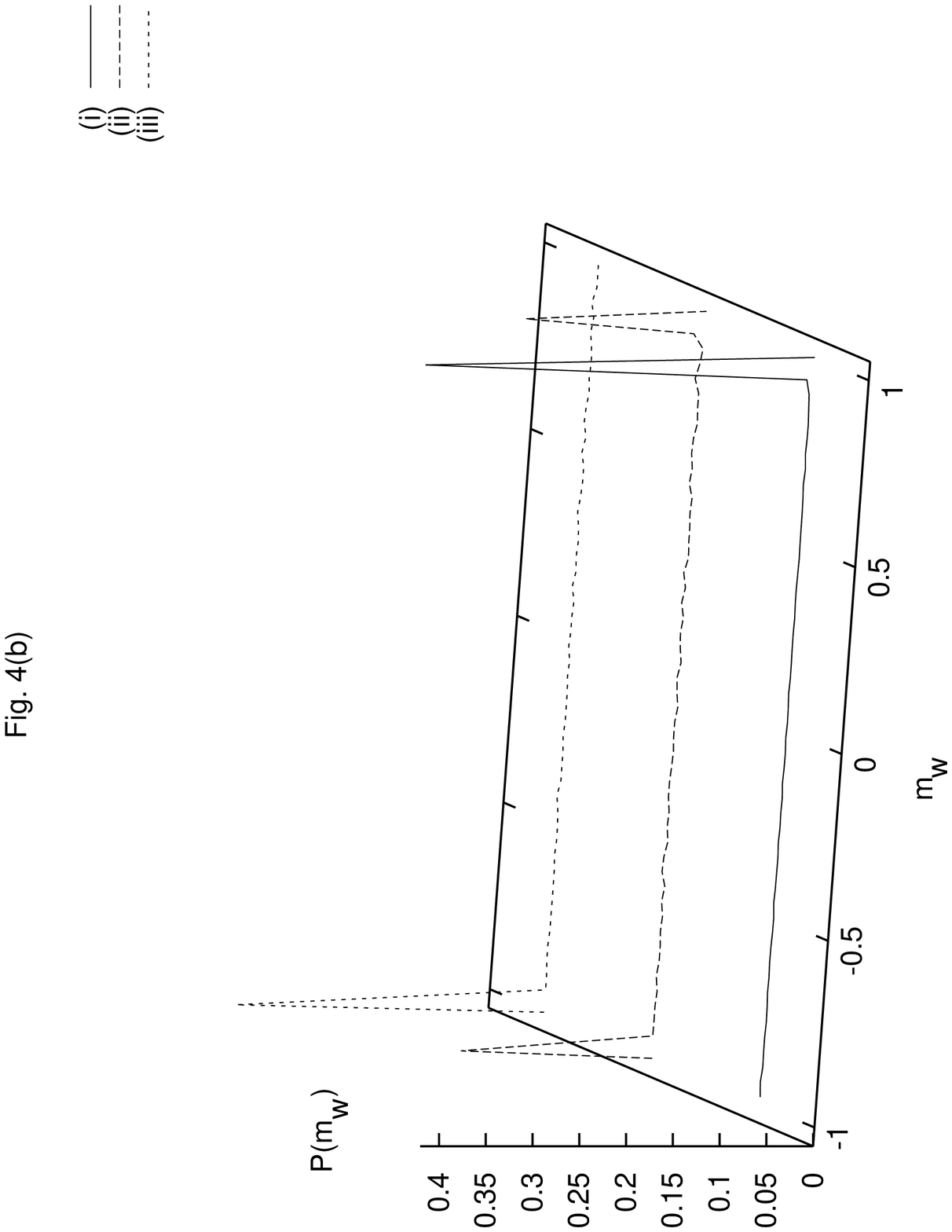} \par}
\vspace{0.3cm}

\end{document}